\documentclass[epj,twocolumn]{webofc}
\usepackage[varg]{txfonts}   
\usepackage{array,amsmath,amssymb,epsfig,fleqn,url,psboxit,color,here,graphics,graphicx,rotating,multirow}
\usepackage{amsfonts,lscape,ulem,nicefrac,mathrsfs,wasysym,hyperref}
\usepackage{pdfsync}
\usepackage[T1]{fontenc}
\usepackage{pbox}
\usepackage{dcolumn}
\usepackage{bigstrut}

\hypersetup{
    colorlinks,
    citecolor=black,
    filecolor=black,
    linkcolor=black,
    urlcolor=black
}

\newcommand{\slim}{\mskip 1.5mu}              

\newcommand{\beq}{\begin{equation}}
\newcommand{\eeq}{\end{equation}}
\newcommand{\beqn}{\begin{eqnarray}}
\newcommand{\eeqn}{\end{eqnarray}}
\newcommand{\bit}{\begin{itemize}}
\newcommand{\eit}{\end{itemize}}
\newcommand{\bef}{\begin{figure}}
\newcommand{\eef}{\end{figure}}
\newcommand{\bec}{\begin{center}}
\newcommand{\eec}{\end{center}}

\woctitle{TRANSVERSITY 2014}
\begin{document}
\title{New results on transverse spin asymmetries from COMPASS}
%
%

\author{Bakur Parsamyan\inst{1,2}\fnsep\thanks{\email{bakur.parsamyan@cern.ch}}
}

\institute{Dipartimento di Fisica Generale, Universit\`a di Torino, Torino, Italy
\and
           INFN, Sezione di Torino, Via P. Giuria 1, I-10125 Torino, Italy
          }

\abstract{%
One of the important objectives of the COMPASS experiment is the exploration of transverse spin structure of nucleon via spin (in)dependent
azimuthal asymmetries in semi-inclusive deep inelastic scattering (SIDIS) of polarized leptons
(and soon also Drell-Yan (DY) reactions with $\pi^-$) off transversely polarized target.
For this purpose a series of measurements were made in COMPASS, using 160 GeV/c longitudinally polarized muon beam and polarized $^6LiD$ and $NH_3$ targets
and are foreseen with 190 GeV/c $\pi^-$ beam on polarized $NH_3$.
	The experimental results obtained by COMPASS for azimuthal effects in SIDIS
play an important role in the general understanding of the three-dimensional nature of the nucleon and are widely used in theoretical analyses and global data fits.
Future first ever polarized DY-data from COMPASS compared with SIDIS results will open a new chapter probing general principles of QCD TMD-formalism.
	In this review main focus will be given to the very recent COMPASS results obtained for SIDIS transverse spin asymmetries from four "Drell-Yan" $Q^2$-ranges.
}
\maketitle
\section{Introduction}
\label{sec:intro}
\begin{figure}
\centering
\label{fig:angles}       
\includegraphics[width=8.0cm,clip]{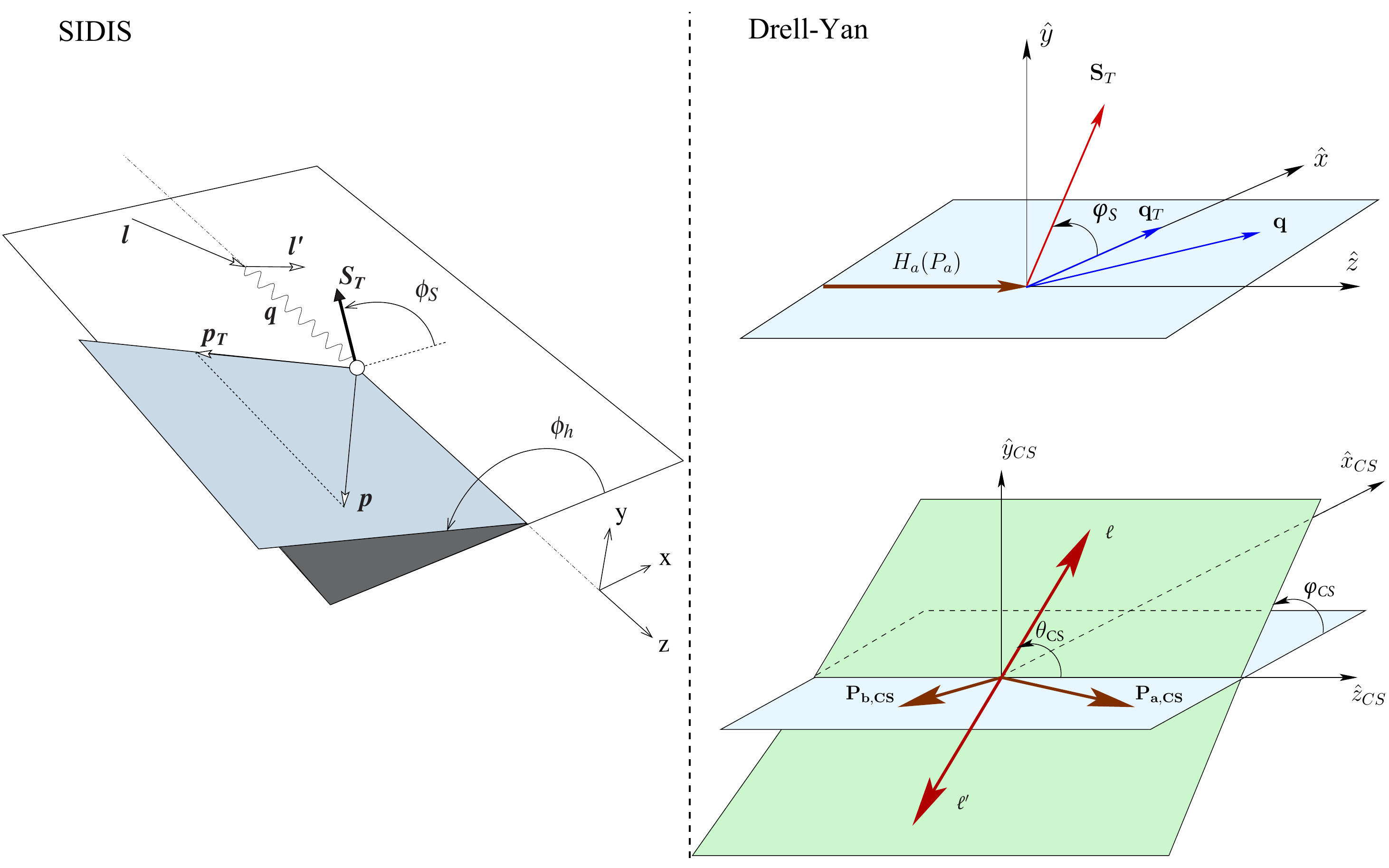}
\caption{SIDIS and Drell-Yan frameworks and notations.}
\end{figure}

Using standard notations SIDIS cross-section can be written in a following model-independent way ~\cite{Kotzinian:1994dv}, \cite{Bacchetta:2006tn}:
{\small
\begin{eqnarray}\label{eq:SIDIS}       
&&\hspace*{-1.6cm}
  \frac{{d\sigma }}{{dxdydzdP_{hT}^2d{\varphi _h}d\psi }} = \left[ {\frac{\alpha }{{xy{Q^2}}}\frac{{{y^2}}}{{2\left( {1 - \varepsilon } \right)}}\left( {1 + \frac{{{\gamma ^2}}}{{2x}}} \right)}\right]F_{UU} \hfill \\ \nonumber
&&\hspace*{-1.5cm}
  \times \left\{ {1 + \sqrt {2\varepsilon \left( {1 + \varepsilon } \right)} \textcolor[rgb]{0.00,0.07,1.00}{A_{UU}^{\cos {\phi _h}}}\cos {\phi _h} + \varepsilon \textcolor[rgb]{1.00,0.00,0.00}{A_{UU}^{\cos 2{\phi _h}}}\cos 2{\phi _h}} \right. \hfill \\ \nonumber
&&\hspace*{-1.4cm}
  {\text{ }} + {\text{ }}\lambda \left[ {\sqrt {2\varepsilon \left( {1 - \varepsilon } \right)} \textcolor[rgb]{0.00,0.07,1.00}{A_{LU}^{\sin {\phi _h}}}\sin {\phi _h}} \right] \hfill \\ \nonumber
&&\hspace*{-1.4cm}
  {\text{ }} + {\text{ }}{S_{L{\text{ }}}}\left[ {\sqrt {2\varepsilon \left( {1 + \varepsilon } \right)} \textcolor[rgb]{0.00,0.07,1.00}{A_{UL}^{\sin {\phi _h}}}\sin {\phi _h} + \varepsilon \textcolor[rgb]{1.00,0.00,0.00}{A_{UL}^{\sin 2{\phi _h}}}\sin 2{\phi _h}} \right] \hfill \\ \nonumber
&&\hspace*{-1.4cm}
  {\text{ }} + {\text{ }}{S_L}\lambda \left[ {\sqrt {1 - {\varepsilon ^2}} \textcolor[rgb]{1.00,0.00,0.00}{{A_{LL}}} + \sqrt {2\varepsilon \left( {1 - \varepsilon } \right)} \textcolor[rgb]{0.00,0.07,1.00}{A_{LL}^{\cos {\phi _h}}}\cos {\phi _h}} \right] \hfill \\ \nonumber
&&\hspace*{-1.4cm}
  {\text{ }} + {\text{ }}{{\text{S}}_{\text{T}}}\left[ \textcolor[rgb]{1.00,0.00,0.00}{{A_{UT}^{\sin \left( {{\phi _h} - {\phi _S}} \right)}}}\sin \left( {{\phi _h} - {\phi _S}} \right) \right. \hfill \\ \nonumber
&&\hspace*{-1.0cm}
  {\text{       }} + {\text{ }}\varepsilon \textcolor[rgb]{1.00,0.00,0.00}{A_{UT}^{\sin \left( {{\phi _h} + {\phi _S}} \right)}}\sin \left( {{\phi _h} + {\phi _S}} \right) \hfill \\ \nonumber
&&\hspace*{-1.0cm}
  {\text{       }} + {\text{ }}\varepsilon \textcolor[rgb]{1.00,0.00,0.00}{A_{UT}^{\sin \left( {3{\phi _h} - {\phi _S}} \right)}}\sin \left( {3{\phi _h} - {\phi _S}} \right) \hfill \\ \nonumber
&&\hspace*{-1.0cm}
  {\text{       }} + {\text{ }}\sqrt {2\varepsilon \left( {1 + \varepsilon } \right)} \textcolor[rgb]{0.00,0.07,1.00}{A_{UT}^{\sin {\phi _S}}}\sin {\phi _S} \hfill \\ \nonumber
&&\hspace*{-1.04cm}
  \left. {{\text{       }} + {\text{ }}\sqrt {2\varepsilon \left( {1 + \varepsilon } \right)} \textcolor[rgb]{0.00,0.07,1.00}{A_{UT}^{\sin \left( {2{\phi _h} - {\phi _S}} \right)}}\sin \left( {2{\phi _h} - {\phi _S}} \right)} \right] \hfill \\ \nonumber
&&\hspace*{-1.4cm}
  {\text{ }} + {\text{ }}{{\text{S}}_{\text{T}}}\lambda \left[ {\sqrt {\left( {1 - {\varepsilon ^2}} \right)} \textcolor[rgb]{1.00,0.00,0.00}{A_{LT}^{\cos \left( {{\phi _h} - {\phi _S}} \right)}}\cos \left( {{\phi _h} - {\phi _S}} \right)} \right. \hfill \\ \nonumber
&&\hspace*{-0.8cm}
  {\text{         }} + {\text{ }}\sqrt {2\varepsilon \left( {1 - \varepsilon } \right)} \textcolor[rgb]{0.00,0.07,1.00}{A_{LT}^{\cos {\phi _S}}}\cos {\phi _S} \hfill \\ \nonumber
&&\hspace*{-0.8cm}
  {\text{         }} + \hspace*{-0.1cm}\left. {\left. {{\text{ }}\sqrt {2\varepsilon \left( {1 - \varepsilon } \right)} \textcolor[rgb]{0.00,0.07,1.00}{A_{LT}^{\cos \left( {2{\phi _h} - {\phi _S}} \right)}}\cos \left( {2{\phi _h} - {\phi _S}} \right)} \right]} \right\}{\text{ }} \hfill \nonumber
\end{eqnarray}
}
where,  $F_{UU}=F_{UU,T}+\varepsilon F_{UU,L}$ and $\psi$ is the laboratory azimuthal angle of the scattered
lepton (in DIS kinematics $d\psi \approx d\phi_S$).
%
 Target transverse polarization dependent part of this general expression contains eight azimuthal modulations in the $\phi_h$
and $\phi_S$ azimuthal angles of the produced hadron and of the nucleon spin, correspondingly, see Fig.~\ref{fig:angles}. Each modulation leads to a $A_{BT}^{w_i(\phi_h, \phi_s)}$
Target-Spin-dependent Asymmetry (TSA) defined as a ratio of the associated structure function to the unpolarized ones. Here the superscript of the asymmetry
indicates corresponding modulation, the first and the second subscripts - respective ("U"-unpolarized,"L"-longitudinal and
"T"-transverse) polarization of beam and target. Five amplitudes which depend only on $S_T$ are the Single-Spin Asymmetries (SSA), the other three
which depend both on $S_T$ and $\lambda$ (beam longitudinal polarization) are known as Double-Spin Asymmetries (DSA).
Amplitude of each modulation is scaled by a $\varepsilon$-dependent so-called depolarization factor where:
\begin{eqnarray}\label{e:epsilon}
&\varepsilon = \frac{1-y -\frac{1}{4}\slim
\gamma^2 y^2}{1-y +\frac{1}{2}\slim y^2 +\frac{1}{4}\slim \gamma^2
y^2}, \,\,\gamma = \frac{2 M x}{Q}
\label{eq:epsilonG}       
\end{eqnarray}
Using similar notations, the general form of the single-polarized ($\pi N^\uparrow$) Drell-Yan cross-section (leading order part) in terms of angular variables defined in
Collins-Soper frame (Fig.~\ref{fig:angles}) can be written in the following 
%
%
model-independent way~\cite{proposal_II}:
{\small
\begin{eqnarray}\label{eq:DY}       
  &&\hspace*{-1.6cm}\frac{{d{\sigma ^{LO}}}}{{d\Omega }} = \frac{{\alpha _{em}^2}}{{F{q^2}}}F_U^1 \left\{ {1 + {{\cos }^2}\theta  + {{\sin }^2}\theta \textcolor[rgb]{1.00,0.00,0.00}{A_U^{\cos 2{\varphi _{CS}}}}\cos 2{\varphi _{CS}}} \right. \hfill \\ \nonumber
  &&\hspace*{+0.5cm}{\text{   }} + {S_L}{\sin ^2}\theta \textcolor[rgb]{1.00,0.00,0.00}{A_L^{\sin 2{\varphi _{CS}}}}\sin 2{\varphi _{CS}} \hfill \\ \nonumber
  &&\hspace*{+0.5cm}{\text{   }} + {S_T}\left[ {\left( {1 + {{\cos }^2}\theta } \right)\textcolor[rgb]{1.00,0.00,0.00}{A_T^{\sin {\varphi _S}}}\sin {\varphi _S}} \right. \hfill \\ \nonumber
  &&\hspace*{+1.2cm}{\text{            }} + {\text{ }}{\sin ^2}\theta \textcolor[rgb]{1.00,0.00,0.00}{A_T^{\sin \left( {2{\varphi _{CS}} + {\varphi _S}} \right)}}\sin \left( {2{\varphi _{CS}} + {\varphi _S}} \right) \hfill \\ \nonumber
  &&\hspace*{+1.2cm}{\text{            }} + \left. {\left. { {{\sin }^2}\theta \textcolor[rgb]{1.00,0.00,0.00}{A_T^{\sin \left( {2{\varphi _{CS}} - {\varphi _S}} \right)}}\sin \left( {2{\varphi _{CS}} - {\varphi _S}} \right)} \right]} \right\} \hfill \nonumber
\end{eqnarray}
}
Similarly to the SIDIS case, the superscript of the asymmetry
indicates the corresponding modulation. As in Eq.~\ref{eq:SIDIS} "U","L" and "T" denote the state of the target polarization.
As one can see, in the Drell-Yan cross-section only one unpolarized and three target transverse spin dependent azimuthal modulations arise at leading order.

Within the QCD parton model approach four of
the eight SIDIS TSAs have Leading Order (LO) interpretation (first three SSAs and first DSA in Eq.~\ref{eq:SIDIS}) and are
described by the different convolutions of Transverse Momentum Dependent (TMD)
twist-two distribution functions (DFs) and fragmentation functions(FFs)
\footnote{In Eq.~\ref{eq:SIDIS}, Eq.~\ref{eq:DY}, Table.~\ref{tab:PDFs} and Table.~\ref{tab:PT2ST} the "twist-2" (LO) amplitudes are marked in red and those which have higher twist interpretation - in blue}.
The first two are well-known Sivers (gives access to "Sivers" PDF $f_{1T}^{\perp\,q}$) and Collins (gives access to "transversity" PDF $h_1^q$)
asymmetries. The other two $A_{UT}^{\sin
(3\phi _h -\phi _s )}$ and $A_{LT}^{\cos (\phi _h -\phi _s )}$ LO TSAs are related to the $h_{1T}^{\perp\,q}$ (pretzelosity) and $g_{1T}^q$ (worm-gear) DFs, correspondingly.
The remaining four SIDIS asymmetries are higher-twist effects, however they
can be expressed in terms of twist-two PDFs being interpreted as Cahn kinematic corrections to twist-two spin
effects on the transversely polarized nucleon (suppressed with respect to the leading twist ones by $\sim M/Q$)
(for details see: \cite{Bacchetta:2006tn},\cite{Mulders:1995dh}-\cite{Parsamyan:2007ju}).

Within the same QCD parton model approach, Drell-Yan TSAs are also interpreted in terms of TMD PDFs.
In this case the asymmetries are related to the convolution of two TMD PDFs: one of the beam and one of the target hadron.
Quoting only the target nucleon PDFs: the $A_{T}^{\sin \varphi _s }$, $A_{T}^{\sin
(2\varphi _{CS} -\varphi _s )}$ and $A_{T}^{\sin(2\varphi _{CS} +\varphi _s )}$ give access to the
"Sivers" $f_{1T}^{\perp\,q}$, "transversity" $h_1^q$ and "pretzelosity" $h_{1T}^{\perp\,q}$, distribution functions, respectively.
Within the concept of \textit{universality} (process-independence) of TMD PDFs it appears that same parton distributions functions can be accessed
both in SIDIS and Drell-Yan (see the Table.~\ref{tab:PDFs} for the complete list).
{\footnotesize
\begin {table}[H]\label{tab:PDFs}
\caption {Nucleon TMD PDFs accessed via SIDIS and Drell-Yan asymmetries}
\begin{center}
\begin{tabular}{ccc}
  \hline
  SIDIS $\ell^\rightarrow N^\uparrow$ & TMD PDF & DY $\pi N^\uparrow$ (LO)\bigstrut\\ \hline
  \textcolor[rgb]{1.00,0.00,0.00}{$A_{UU}^{\cos 2\phi _h}$}, \textcolor[rgb]{0.00,0.00,1.00}{$A_{UU}^{\cos \phi _h}$} & $h_{1}^{\bot q}$& \textcolor[rgb]{1.00,0.00,0.00}{$A_{U}^{\cos 2\varphi _{CS}}$} \bigstrut\\ \hline
  \textcolor[rgb]{1.00,0.00,0.00}{$A_{UT}^{\sin (\phi _h -\phi _s )}$}, \textcolor[rgb]{0.00,0.00,1.00}{$A_{UT}^{\sin\phi _s}$}, \textcolor[rgb]{0.00,0.00,1.00}{$A_{UT}^{\sin (2\phi _h -\phi _s )}$} & $f_{1T}^{\bot q}$& \textcolor[rgb]{1.00,0.00,0.00}{$A_{T}^{\sin \varphi _{S}}$} \bigstrut\\ \hline
  \textcolor[rgb]{1.00,0.00,0.00}{$A_{UT}^{\sin (\phi _h +\phi _s -\pi)}$}, \textcolor[rgb]{0.00,0.00,1.00}{$A_{UT}^{\sin\phi _s}$} & $h_{1}^{q}$& \textcolor[rgb]{1.00,0.00,0.00}{$A_{T}^{\sin (2\varphi _{CS} -\varphi _{S} )}$} \bigstrut\\ \hline
  \textcolor[rgb]{1.00,0.00,0.00}{$A_{UT}^{\sin (3\phi _h -\phi _s )}$}, \textcolor[rgb]{0.00,0.00,1.00}{$A_{UT}^{\sin (2\phi _h -\phi _s )}$} & $h_{1T}^{\bot q}$& \textcolor[rgb]{1.00,0.00,0.00}{$A_{T}^{\sin (2\varphi _{CS} +\varphi _{S} )}$} \bigstrut\\ \hline
  \textcolor[rgb]{1.00,0.00,0.00}{$A_{LT}^{\cos (\phi _h -\phi _s )}$}, \textcolor[rgb]{0.00,0.00,1.00}{$A_{LT}^{\cos\phi _s}$}, \textcolor[rgb]{0.00,0.00,1.00}{$A_{LT}^{\cos (2\phi _h -\phi _s )}$} & $g_{1T}^{ q}$& \textcolor[rgb]{1.00,0.00,0.00}{DP DY} \bigstrut\\ \hline
\end{tabular}
\end{center}
\end {table}
}
Therefore, DY measurements at COMPASS will be intriguingly complementary to the COMPASS SIDIS results and
will give an unprecedented opportunity to access TMD PDFs via two mechanisms and test their universality and key features
(for instance, predicted Sivers and Boer-Mulders PDFs sign change) using essentially same experimental setup.
Certainly, at some point both sets of COMPASS results from SIDIS and Drell-yan will be a subject of global fits and phenomenological comparison.
For this purpose the best option is to explore SIDIS data in more differential way extracting the asymmetries in the same four $Q^2$
kinematic regions (which implies also different $x$-coverage) which were selected for the COMPASS Drell-Yan measurement program ~\cite{proposal_II}:
\begin{eqnarray}\label{eq:DY_ranges}
    &&\hspace*{-1.4cm}\bullet\hspace*{0.3cm} 1<Q^{2}/(GeV/c)^2<4\ \ "low\ mass"\\ \nonumber
    &&\hspace*{-1.4cm}\bullet\hspace*{0.3cm} 4<Q^{2}/(GeV/c)^2<6.25\ \ "intermediate\ mass"\\ \nonumber
    &&\hspace*{-1.4cm}\bullet\hspace*{0.3cm} 6.25<Q^{2}/(GeV/c)^2<16\ \ "J/\psi\ mass"\\ \nonumber
    &&\hspace*{-1.4cm}\bullet\hspace*{0.3cm} Q^{2}/(GeV/c)^2>16\ \ "High\ mass". \nonumber
\end{eqnarray}
Here the most promising is the so-called "high mass" range which is expected to be free from background and corresponds to the valence-quark region
where the Drell-Yan asymmetries are expected to reach their largest values \cite{proposal_II}.
SIDIS TSAs extracted from aforementioned "$Q^2$-ranges" will serve not only for future SIDIS-DY comparison,
but, exploring two-dimensional $x$:$Q^2$-behaviour of the asymmetries, they can be used also as a better input
for TMD-evolution studies and related SIDIS-DY predictions \cite{Echevarria:2014xaa},\cite{Sun:2013hua}.
In this review COMPASS results for all SIDIS TSAs extracted from four Drell-Yan $Q^2$-ranges will be discussed.
\begin{figure}[h!]
\centering
\includegraphics[width=8.0cm,clip]{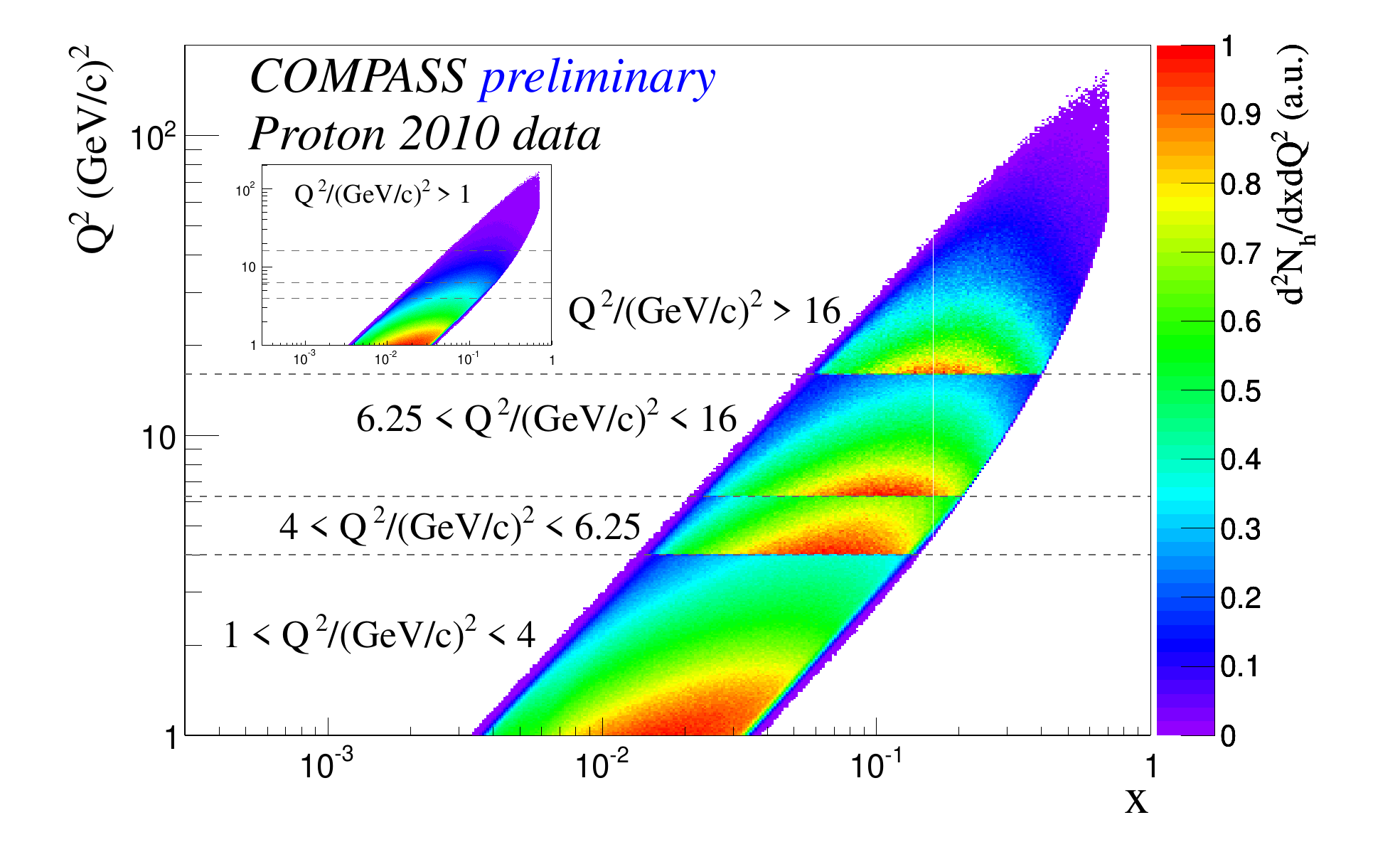}
\includegraphics[width=8.0cm,clip]{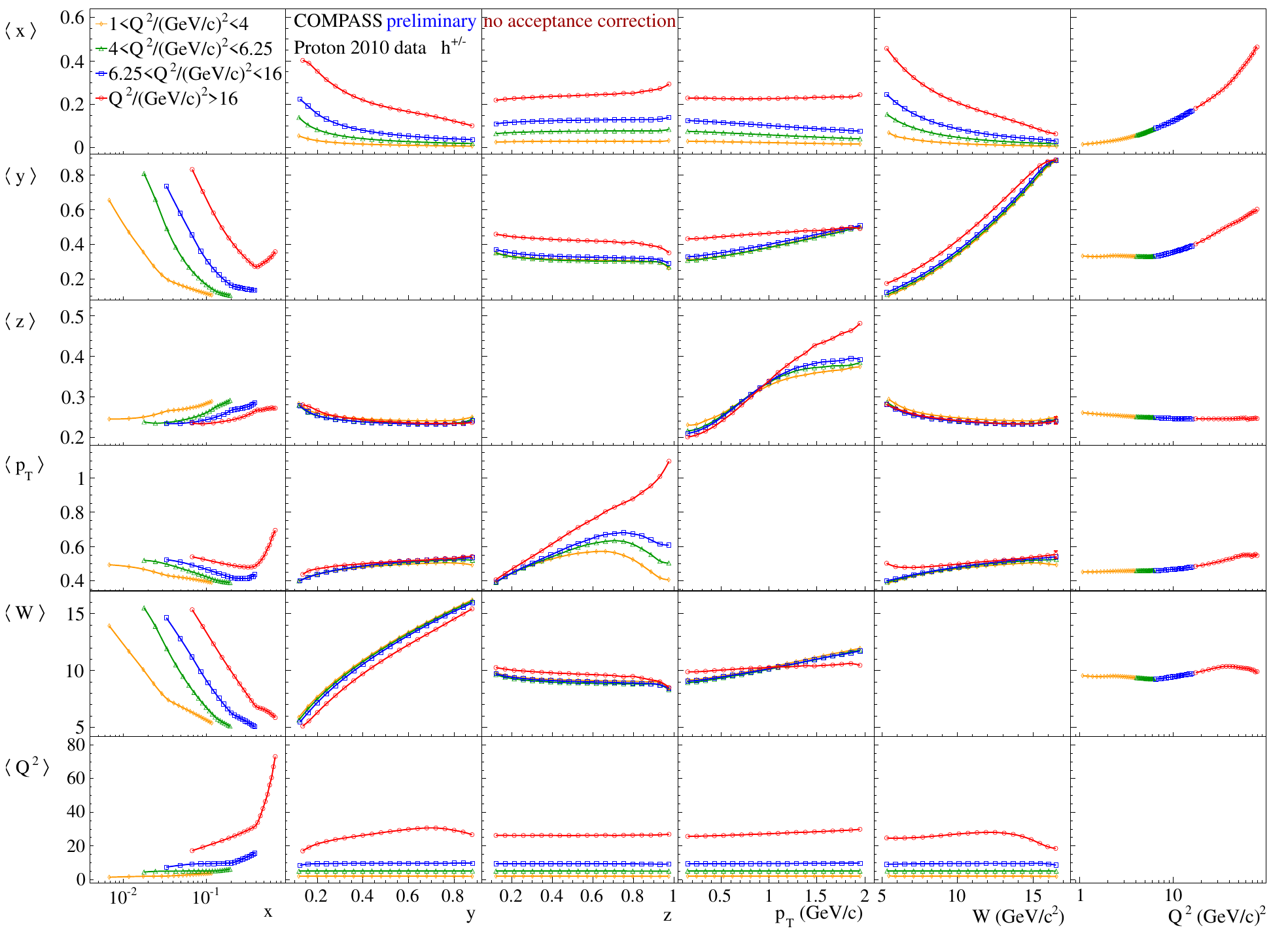}
\caption{COMPASS x$:Q^2$ phase-space with indicated four Drell-Yan $Q^2$-ranges (top). COMPASS multidimensional kinematical "map" (bottom).}
\label{fig:kvar}       
\end{figure}

\section{Data analysis details}
\label{sec:data_analysis}

Asymmetries were extracted from COMPASS 2010 - transversely polarized proton data.
In general, event selection procedure as well as asymmetry extraction and systematic uncertainty definition techniques applied for this analysis
are identical to those used for recent COMPASS results on Collins, Sivers and other TSAs \cite{Adolph:2012sn}-\cite{Parsamyan:2007ju}.

The DIS events are selected by applying standard cuts: $Q^2>1$ $(GeV/c)^2$, $0.003<x<0.7$ and $0.1 <y < 0.9$.
Two more cuts were applied on \textit{hadronic} variables: $p_T>0.1$ GeV/c and $z>0.2$.
In COMPASS kinematics "$0.1<z<0.2$" and "$z>0.2$" data samples contain nearly same number of events and
in order to improve the statistical accuracy and clarify the trends, TSAs have also been studied in extended "$z>0.1$" range.

The $x$:$Q^2$ phase-space was divided into four sub-ranges by selecting four Drell-Yan $Q^2$-bins Fig.~\ref{fig:kvar} (top) according to Eq.~\ref{eq:DY_ranges}.
Multidimensional map of COMPASS kinematical dependencies as extracted from the data is presented in Fig.~\ref{fig:kvar} (bottom).

All TSAs were measured as functions of
$x$, $z$, $p_T$ and $W$ ($W$-dependencies are omitted in this review for brevity) both for positive and negative hadrons.
\begin{figure}[h!]
\centering
\includegraphics[width=8.0cm,clip]{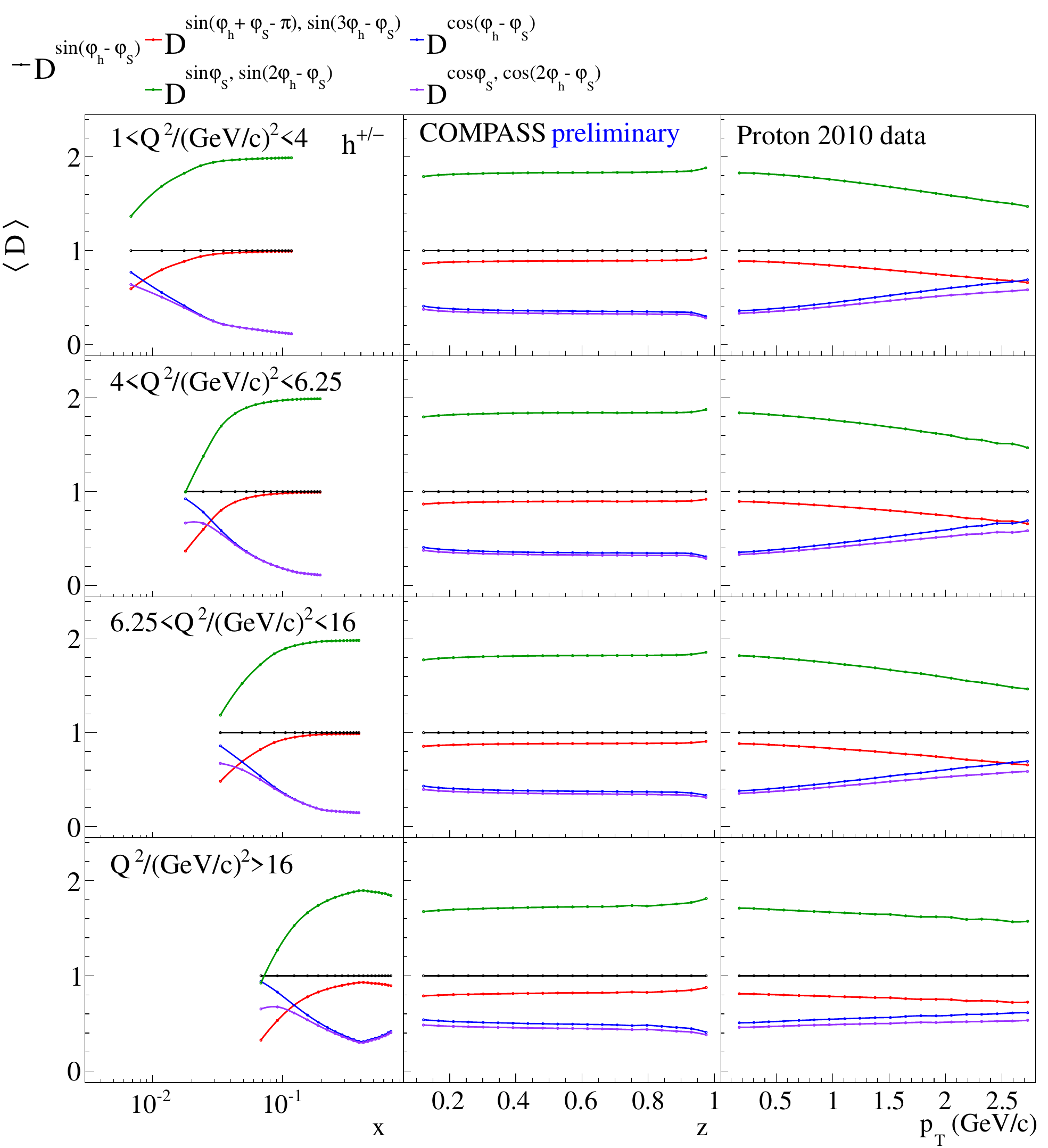}
\caption{Mean depolarization factors.}
\label{fig:Dy}       
\end{figure}
In accordance with Eq.~\ref{eq:SIDIS} "physics" asymmetries for the given modulation $w(\phi_h, \phi_s)$ are
related with the "raw" ones obtained from the fit (as amplitudes of the
corresponding azimuthal modulations), through the following relations:
$A_{UT}^{w(\phi_h, \phi_s)} =\frac{A_{UT, \; raw}^{w(\phi_h, \phi_s)}}{f |P_T| D^{w(\phi_h, \phi_s)}(y)}$ and $A_{LT}^{w(\phi_h, \phi_s)} =\frac{A_{LT, \; raw}^{w(\phi_h, \phi_s)}}{f
\lambda |P_T| D^{w(\phi_h, \phi_s)}(y)}$
where $P_T$,  $f$ and $D^{w(\phi_h, \phi_s)}(y)$ are the mean values (extracted from the data in the given kinematical bin)
of the transverse polarization of the target (w.r.t beam axis), target polarization dilution factor and
of the corresponding depolarization factor (Fig.~\ref{fig:Dy}).

Last important detail was already described in previous COMPASS reviews \cite{Parsamyan:2013fia},\cite{Parsamyan:2013ug} and here will be addressed only briefly.
In the  Eq.~\ref{eq:SIDIS} target transverse polarization ($S_T$) is defined relative to the virtual photon momentum direction
(most natural basis from the theory point of view) while, in experiment transverse polarization of the target
is defined relative to the beam (incoming lepton) direction. As it was demonstrated in \cite{Kotzinian:1994dv},\cite{Diehl:2005pc}
this difference, in particular, influences azimuthal distributions in the final state. In the appropriately modified expression
for the SIDIS cross-section for transversely (w.r.t. lepton beam) polarized target \cite{Parsamyan:2013fia},\cite{Parsamyan:2013ug}
one can find new $sin\theta$-scaled terms and $\theta$-dependent factors
($\theta$ is the angle between $\gamma^*$-direction and initial lepton momenta in lab. frame) and see that
some TSAs are getting mixed up with longitudinal spin asymmetries (LSA).
Anyway, since $\theta$ is rather small in COMPASS kinematics the influence of the additional terms and factors can be neglected in most of the cases.
Essentially, one can can derive following relation between the correct TSAs and those extracted from the fit using "Eq.~\ref{eq:SIDIS}"-approach
and therefore mixed with specific LSAs because of the $\gamma^* p\rightarrow lp$ transition: ${A_T} \approx {A_{T,fit}} - C\left( \varepsilon  \right){A_L}$.
Mixed TSAs and LSAs and corresponding $C(\varepsilon,\theta)$-factors are presented in Table.~\ref{tab:PT2ST}

\begin{figure}
\centering
\includegraphics[width=8.0cm,clip]{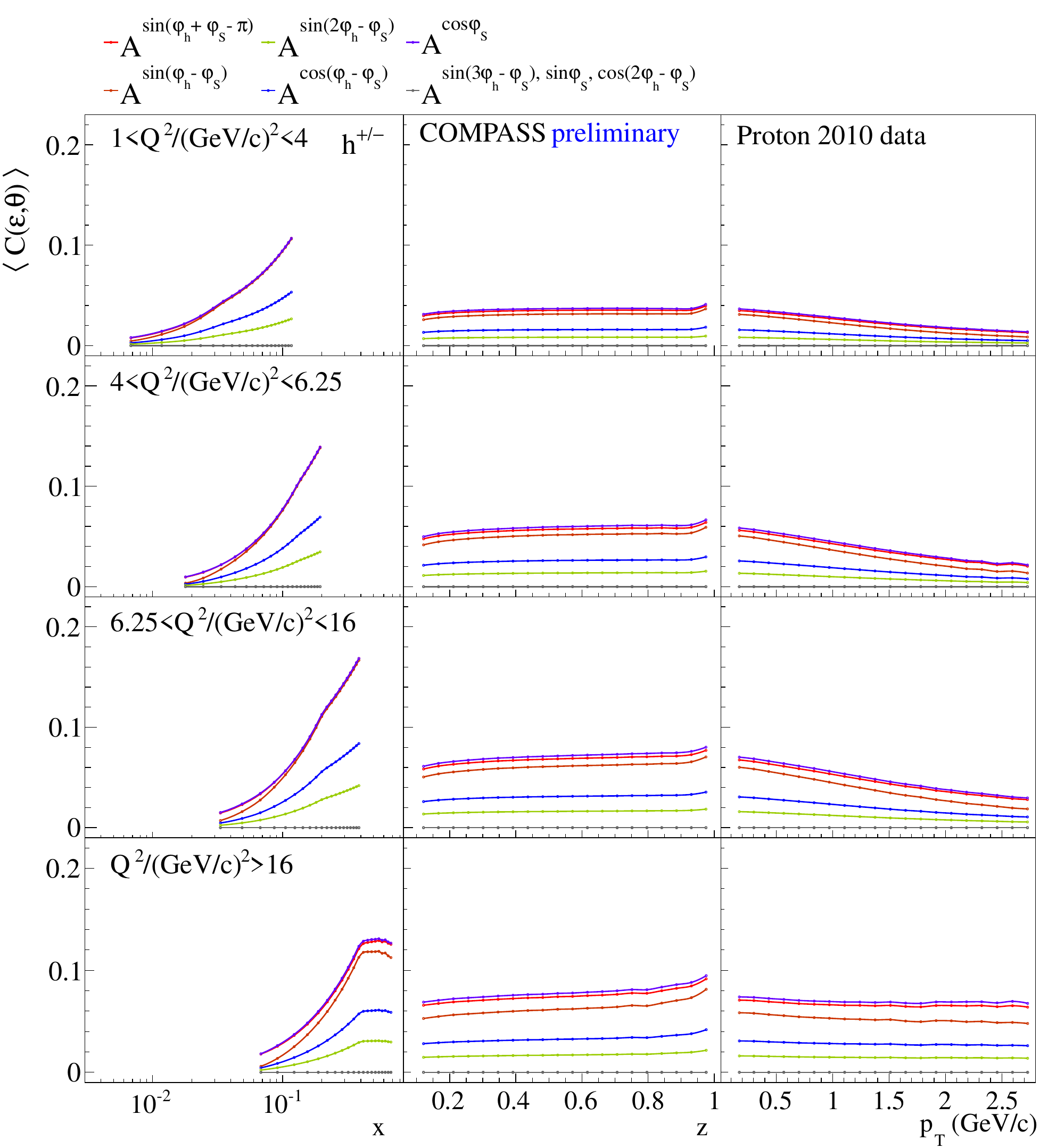}
\includegraphics[width=8.0cm,clip]{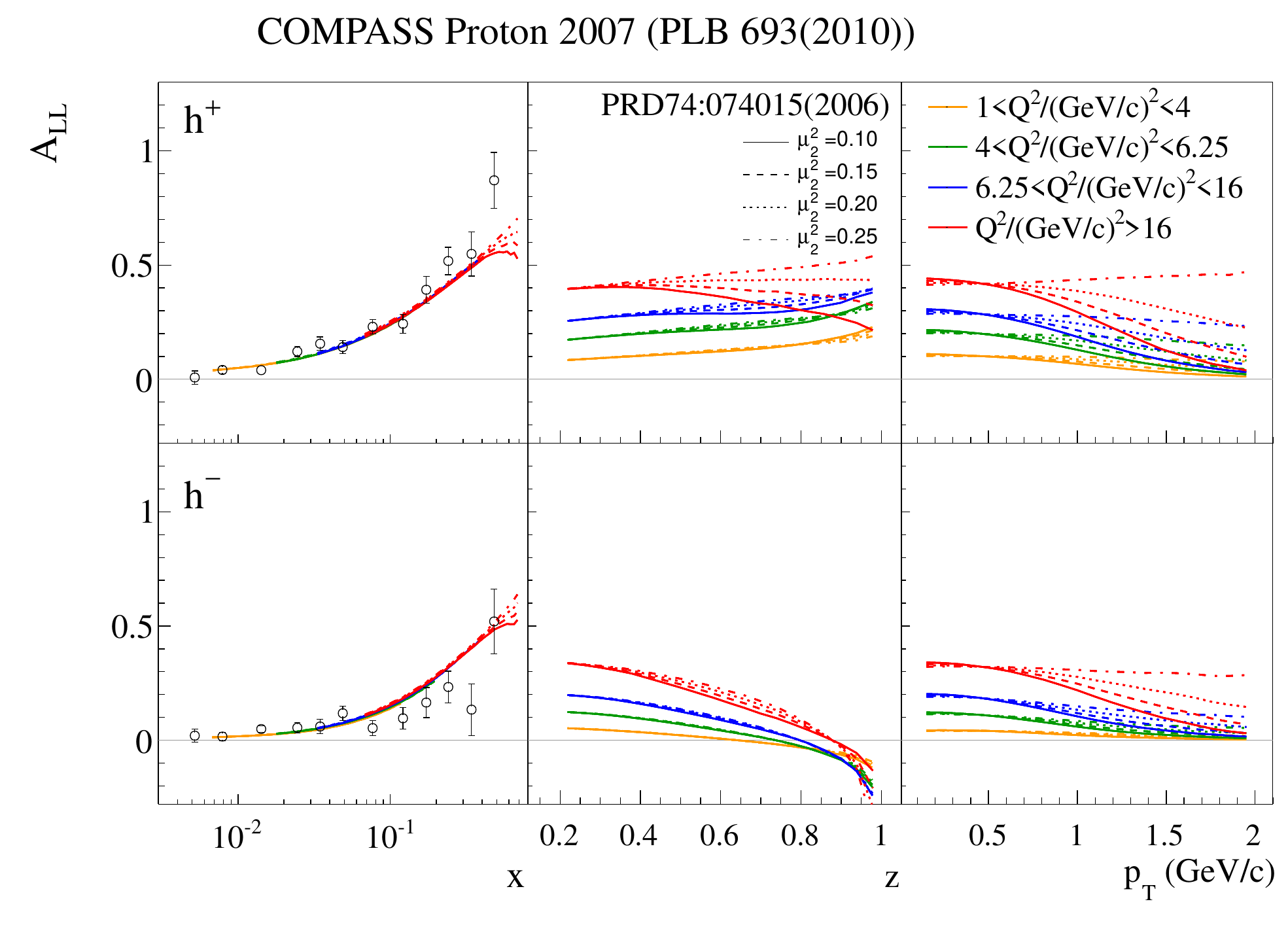}
\caption{Top: mean $C(\varepsilon,\theta)$-factors from Table.~\ref{tab:PT2ST}.
Bottom: $A_{LL}$ asymmetry, COMPASS data ~\cite{Alekseev:2010ub} and predictions ~\cite{Anselmino:2006yc}.}
\label{fig:A_LL}       
\end{figure}

\begin {table}[h!]
\caption {Mixed "T" and "L" amplitudes and $C(\varepsilon,\theta)$-factors}
\begin{center}
\begin{tabular}{ccc}\label{tab:PT2ST}
  TSA & $C(\varepsilon,\theta)$-factor & LSA \bigstrut\\ \hline
  $\textcolor[rgb]{1.00,0.00,0.00}{A_{UT}^{\sin (\phi _h -\phi _s )}}$, $\textcolor[rgb]{1.00,0.00,0.00}{A_{UT}^{\sin (\phi _h +\phi _s -\pi )}}$ & $\frac{ sin\theta\sqrt {2\varepsilon \left( {1 + \varepsilon } \right)}}{2}$ & $\textcolor[rgb]{0.00,0.07,1.00}{A_{UL}^{\sin \phi _h }}$ \bigstrut\\ \hline
  $\textcolor[rgb]{0.00,0.07,1.00}{A_{UT}^{\sin (2\phi _h -\phi _s )}}$ & $\frac{ sin\theta\varepsilon}{2\sqrt {2\varepsilon \left( {1 + \varepsilon } \right)}}$ & $\textcolor[rgb]{1.00,0.00,0.00}{A_{UL}^{\sin 2\phi _h }}$ \bigstrut\\ \hline
  $\textcolor[rgb]{1.00,0.00,0.00}{A_{LT}^{\cos (\phi _h -\phi _s )}}$ & $\frac{ sin\theta\sqrt {2\varepsilon \left( {1 - \varepsilon } \right)}}{2\sqrt {\left( {1 - {\varepsilon ^2}} \right)}}$ & $\textcolor[rgb]{0.00,0.07,1.00}{A_{LL}^{\cos \phi _h }}$ \bigstrut\\ \hline
  $\textcolor[rgb]{0.00,0.07,1.00}{A_{LT}^{\cos \phi _s }}$ & $\frac{ sin\theta\sqrt {\left( {1 - {\varepsilon ^2}} \right)}}{\sqrt {2\varepsilon \left( {1 - \varepsilon } \right)}}$ & $\textcolor[rgb]{1.00,0.00,0.00}{A_{LL}}$ \bigstrut\\ \hline
\end{tabular}
\end{center}
\end {table}
%
%
%
It can be demonstrated that for all transverse asymmetries except $A_{LT}^{\cos {\varphi _S}}$ DSA the impact of TSA-LSA mixing can be neglected.
This is justified by the smallness of $C(\varepsilon,\theta)$-factors in COMPASS kinematics (Fig.\ref{fig:A_LL} (top))
and since also contributing LSAs are measured (or estimated) to be small \cite{Alekseev:2010dm}.
The case of $A_{LT}^{\cos {\varphi _S}}$ asymmetry is peculiar since it is affected by a large $A_{LL}$ amplitude \cite{Alekseev:2010ub}
and here mixing effect cannot be neglected.
In order to correct the $A_{LT}^{\cos {\varphi _S}}$ asymmetry the $A_{LL}$ values evaluated in accordance with \cite{Anselmino:2006yc} were used.
Corresponding $A_{LL}$ curves are shown in Fig.\ref{fig:A_LL} (bottom) compared with the COMPASS data points from \cite{Alekseev:2010ub},
demonstrating close agreement.
\section{Results}
\label{sec:results}
\subsection{Sivers asymmetry}
\label{sec:Sivers}
\begin{figure}[h]
\centering
\includegraphics[width=8.0cm,clip]{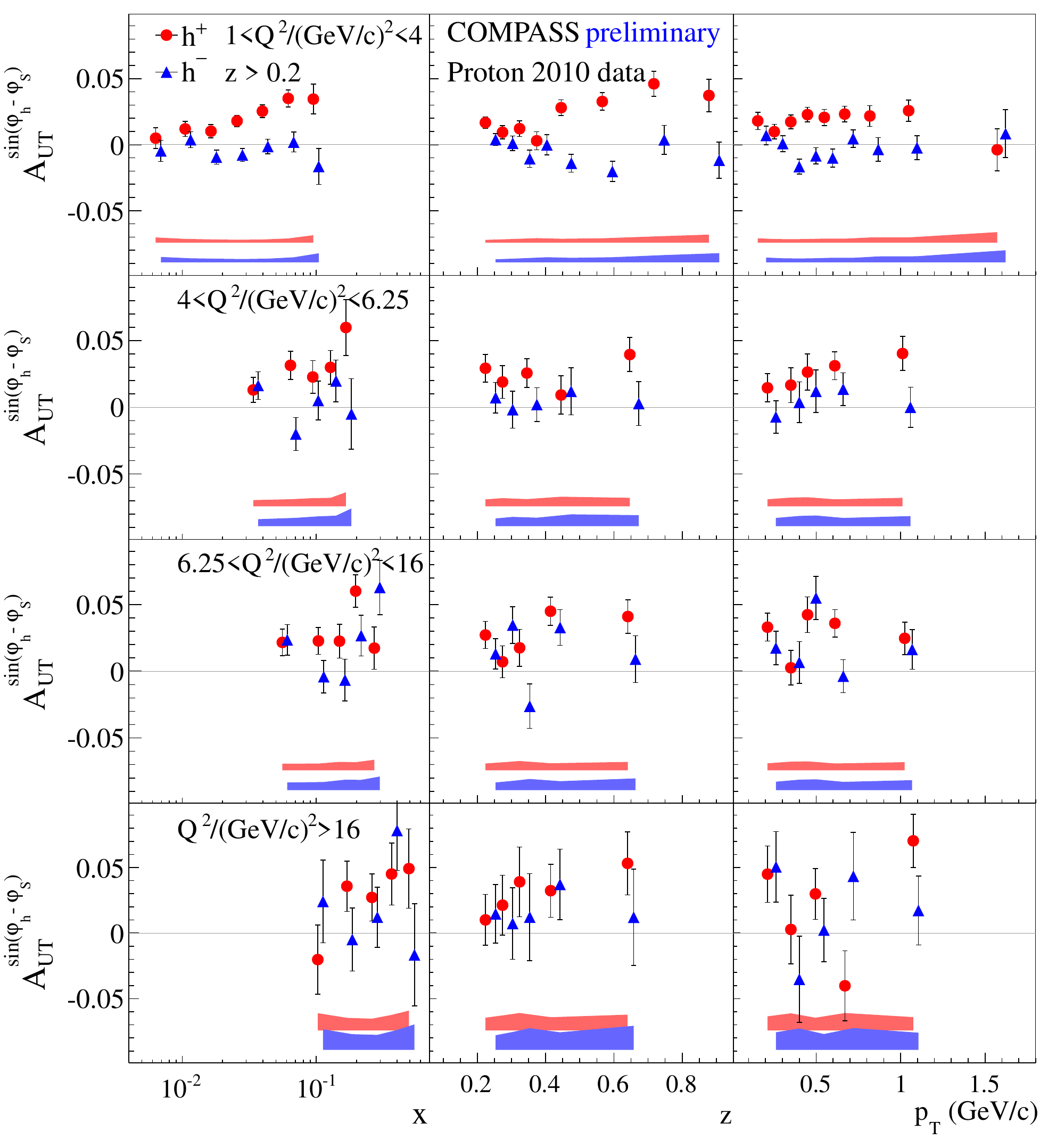}
\includegraphics[width=8.0cm,clip]{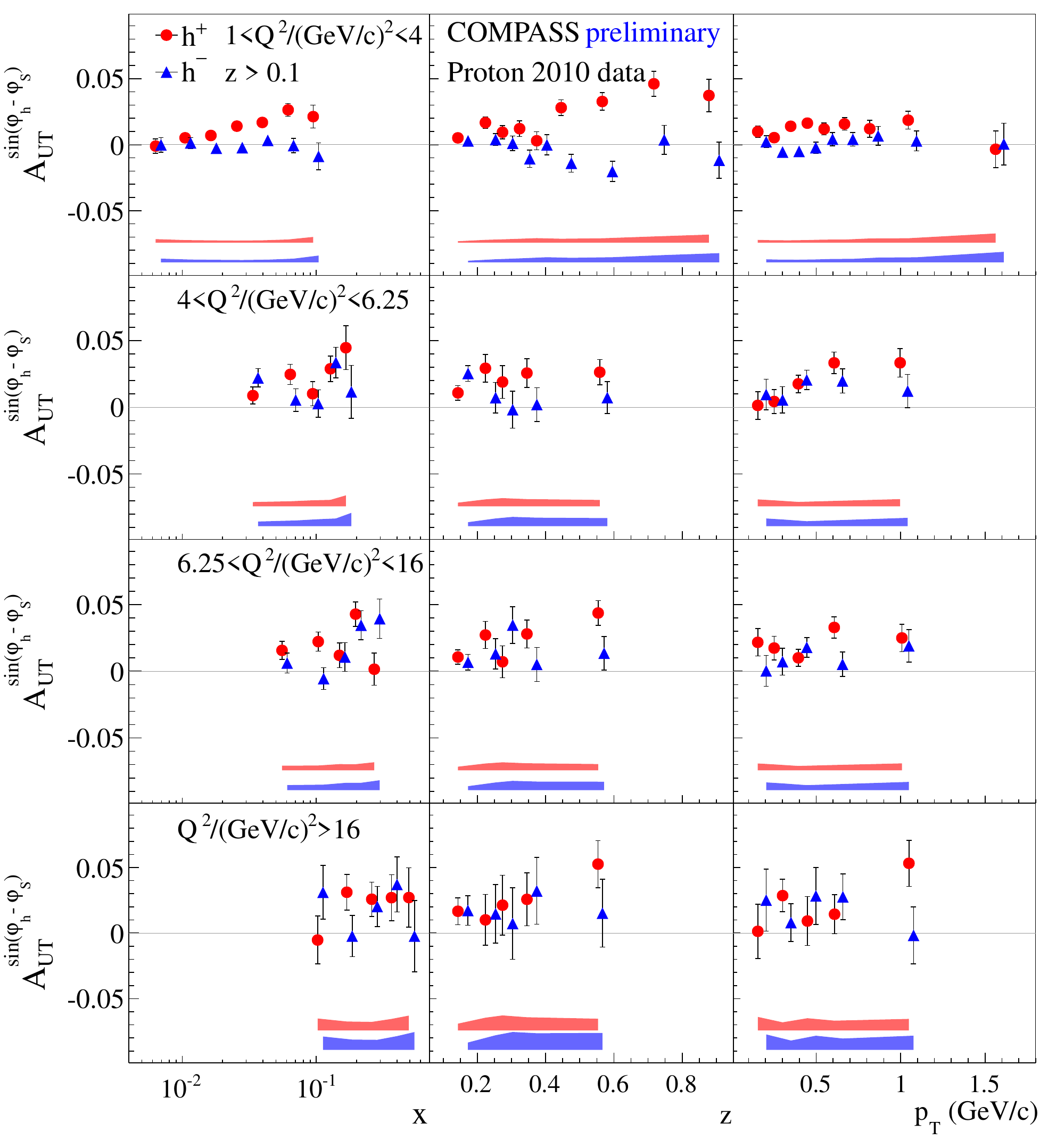}
\caption{Sivers asymmetry: $z>0.2$ (top), $z>0.1$ (bottom)}
\label{fig:Siv}       
\end{figure}

Results for Sivers asymmetry from $z>0.1$ and $z>0.2$ are presented in Fig.~\ref{fig:Siv}.
A clear positive signal is observed for positive hadrons (growing with $x$, $z$ and $p_T$).
For negative hadrons some hints of a negative amplitude can be seen at lowest $Q^2$-range for intermediate $z$ values
while at relatively large $x$ and $Q^2$ there are indication for a positive signal.
Asymmetry appears to be slightly smaller for $z>0.1$ compared to $z>0.2$. Comparing points from same $x$-bins, but different $Q^2$-ranges
one can see that within statistical accuracy there's no clear and strong $Q^2$-dependence for the effect.
Nevertheless, decreasing with $Q^2$ trend can be noted in some bins.
\subsection{Collins asymmetry}
\label{sec:Collins}
For Collins effect, clear signal is visible both for positive and negative hadrons (but with opposite sign) at relatively large $x$ values Fig.~\ref{fig:A_UT1} (top).
Asymmetry grows with $x$, $z$ and $p_T$, but with some "instabilities" (see for instance saddle-shaped trends in two middle $Q^2$-ranges).
No clear $Q^2$ dependence was observed.
\subsection{$A_{LT}^{cos(\phi_h-\phi_S)}$ asymmetry}
\label{sec:A_LT}
The $A_{LT}^{\cos (\phi _h -\phi _s )}$ is the only leading-twist LT-amplitude. It provides access
to $g_{1T}^q(x,k_T^2)$ "worm gear" PDF, which describes longitudinal
polarization of quarks in transversely polarized nucleon. In the Fig.~\ref{fig:A_LT1} results for this asymmetry are shown together with prediction curves
evaluated in accordance with
\cite{Kotzinian:2006dw}. A clear signal is detected for positive and negative hadrons at large $x$ and $Q^2$ values.
Within given statistical accuracy predictions are in agreement with experimental points.
\subsection{$A_{UT}^{sin(\phi_S)}$ asymmetry}
\label{sec:A_UT1}
\begin{figure}[h!]
\centering
\includegraphics[width=8.0cm,clip]{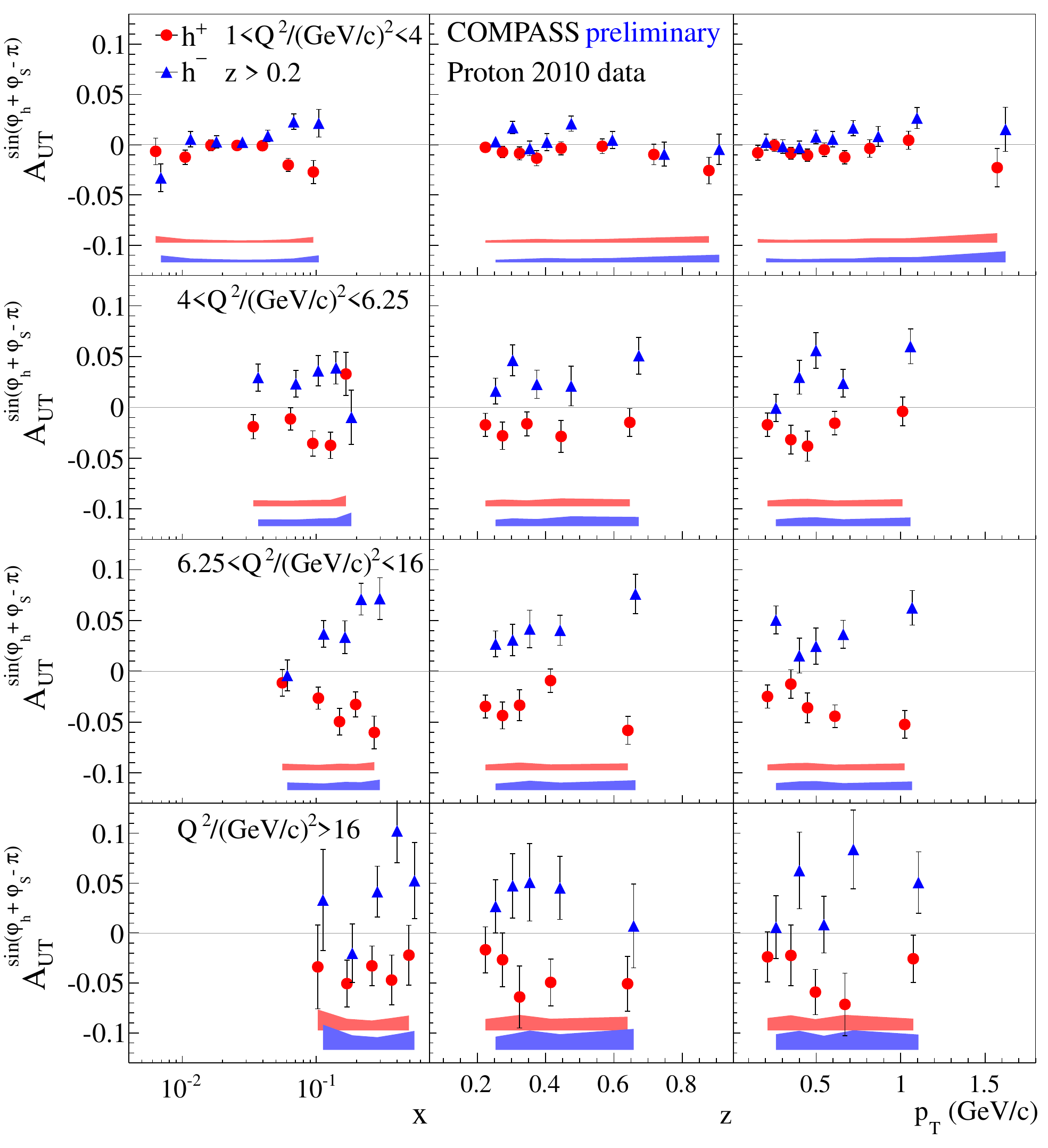}
\includegraphics[width=8.0cm,clip]{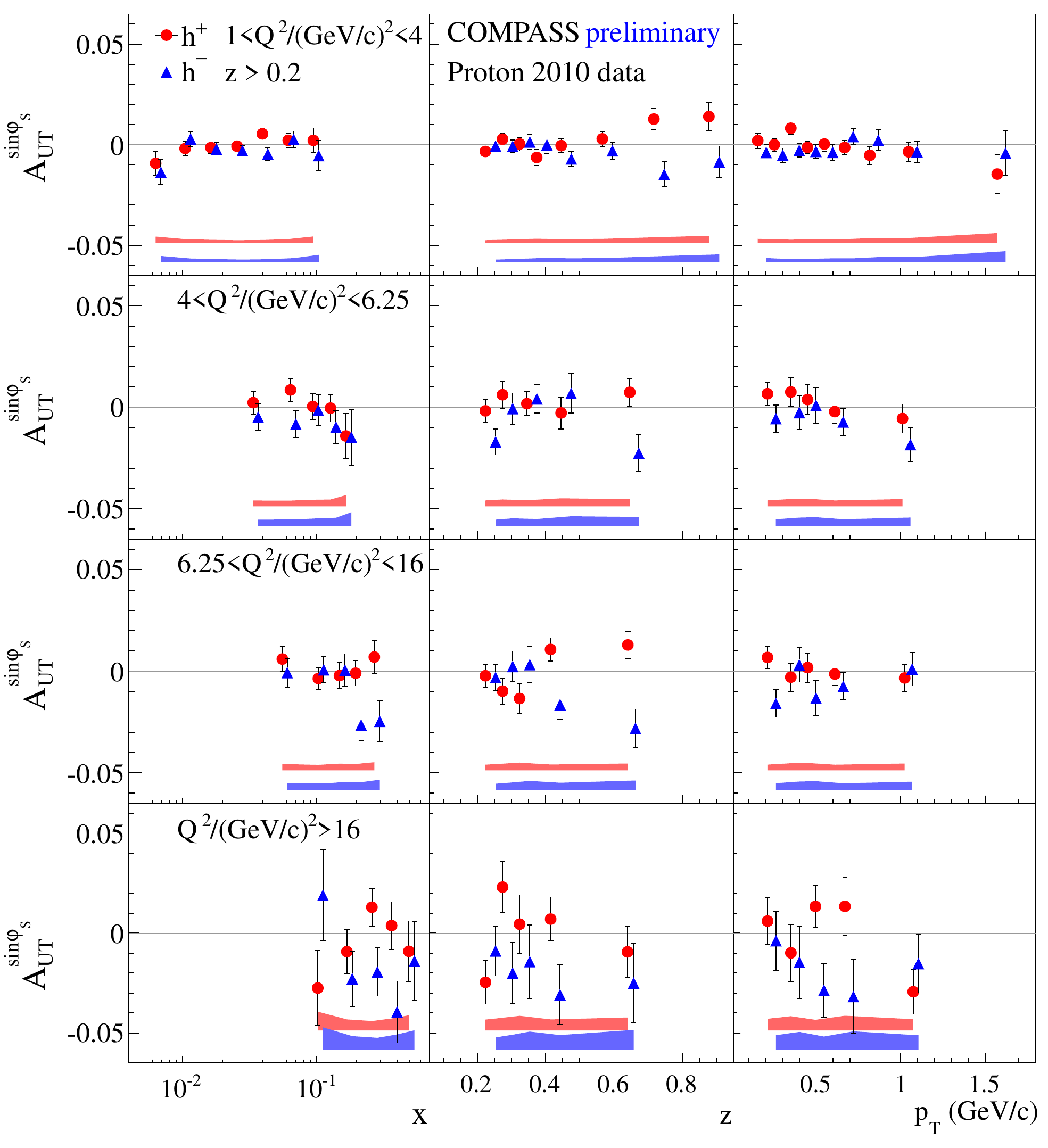}
\caption{Collins (top) and $A_{UT}^{\sin\phi_S}$ (bottom) asymmetries}
\label{fig:A_UT1}       
\end{figure}

The $A_{UT}^{\sin (\phi _s )}$ asymmetry is a sub-leading twist effect.
At first order it can be described by Collins and Sivers mechanisms only,
but suppressed by a factor of $Q^{-1}$ and by a factor of $\sim|{\bf p}_{T}|$ with respect to them. This is the only "higher-twist" effect
which shows non-zero trends. Results for this asymmetry are presented in Fig.~\ref{fig:A_UT1} (bottom).
There are several bins at relatively large $x$ and $Q^2$ values where a negative signal can be seen for negative hadrons. For positive hadrons asymmetry
appears to be small and compatible with zero everywhere, except few large $z$ bins at $Q^2<4 (GeV/c)^2$.
\subsection{$A_{UT}^{sin(3\phi_h-\phi_S)}$, $A_{UT}^{sin(2\phi_h-\phi_S)}$, $A_{LT}^{cos(\phi_S)}$ and $A_{LT}^{cos(2\phi_h-\phi_S)}$ asymmetries}
\label{sec:A_UT1}
\begin{figure}[h!]
\centering
\includegraphics[width=8.0cm,clip]{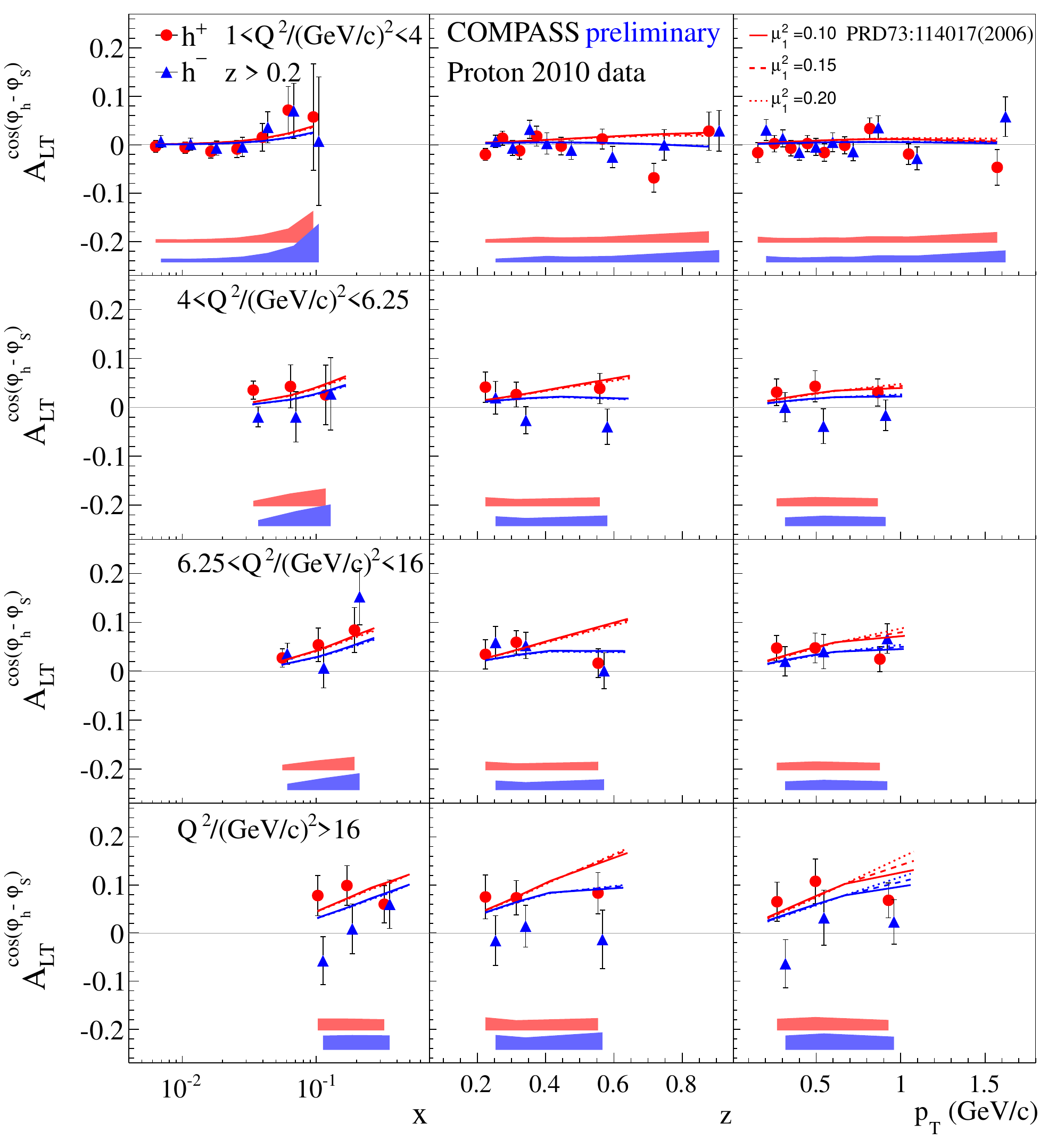}
\includegraphics[width=8.0cm,clip]{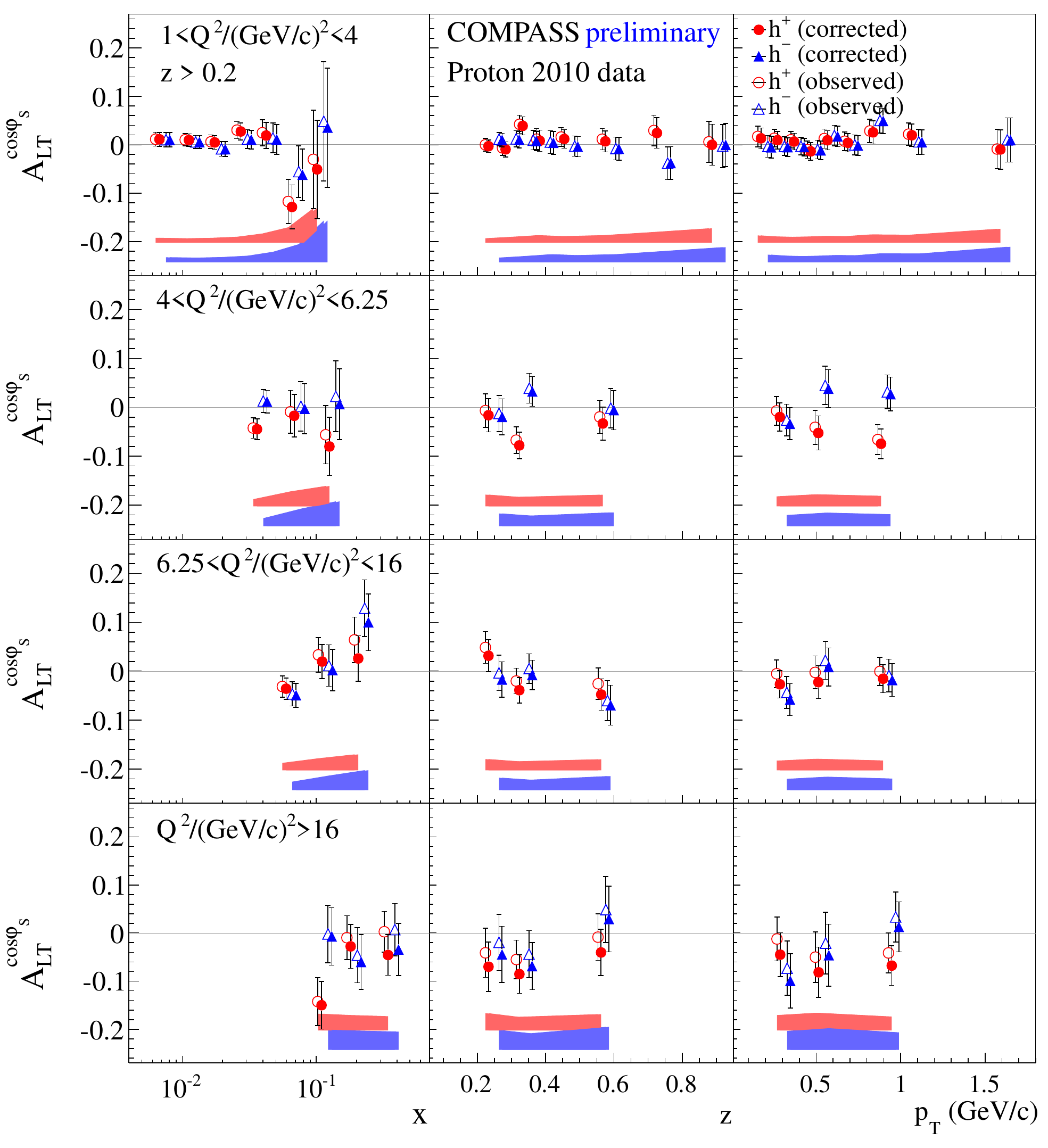}
\caption{Top: $A_{LT}^{\cos(\phi_h-\phi_S)}$ asymmetry and predictions \cite{Kotzinian:2006dw}.
Bottom: $A_{LT}^{\cos\phi_S}$ asymmetry as extracted from the fit and one corrected for $A_{LL}$-contribution ($A_{LL}$ i.a.w. \cite{Anselmino:2006yc}).}
\label{fig:A_LT1}       
\end{figure}

Remaining four asymmetries are found to be compatible with zero within statistical accuracy. This can be explained by different kinematical
suppressions to which they are affected. For instance, the $A_{UT}^{sin(2\phi_h-\phi_S)}$, $A_{LT}^{cos(\phi_S)}$ and $A_{LT}^{cos(2\phi_h-\phi_S)}$
"higher-twist" terms have $Q^{-1}$-suppression \cite{Kotzinian:1994dv}, \cite{Bacchetta:2006tn}, while $A_{UT}^{sin(3\phi_h-\phi_S)}$ leading order amplitude is suppressed by a $\sim|{\bf p}_{T}|^3$ scale-factor \cite{Kotzinian:1994dv}, \cite{Bacchetta:2006tn},\cite{Lefky:2014eia}.
In Fig.~\ref{fig:A_LT1} data-points for the $A_{LT}^{cos(\phi_S)}$ asymmetry are shown as extracted from the fit compared with points corrected for $A_{LL}$-mixing
(see Sec.~\ref{sec:data_analysis}).
\section{Conclusions}
\label{sec:conclusions}
COMPASS provided first input for future direct SIDIS-DY studies. All eight SIDIS TSAs were extracted from four $Q^2$-ranges
selected for the COMPASS future Drell-Yan program and two $z$-selections using proton 2010 transverse data.
Sizable effects were observed for Sivers and Collins amplitudes and clear indications of non-zero asymmetries were taken for the
$A_{LT}^{cos(\phi_h-\phi_S)}$ and $A_{UT}^{sin(\phi_S)}$ asymmetries.
Other four asymmetries were found to be compatible with zero
within given statistical accuracy.
These results combined with future first ever polarized Drell-Yan data from COMPASS will give a
unique opportunity to access TMD PDFs via two processes and test their universality and key features sticking to the same $x$:$Q^2$ kinematical range.
%
%
%

%

\end{document}